\documentstyle[aps,pra,multicol,psfig]{revtex}

\newcommand{\QKD}{{\sc qkd}}

\newcommand{\WCP}{{\sc wcp}}
\newcommand{\PNS}{{\sc pns}}
\newcommand{\QND}{{\sc qnd}}
\newcommand{\CBS}{{\sc cbs}}
\newcommand{\ACBS}{{\sc acbs}}
\newcommand{\DCBS}{{\sc dcbs}}
\newcommand{\CBSF}{{\sc cbsf}}
\newcommand{\BS}{{\sc bs}}
\newcommand{\DBS}{{\sc dbs}}
\newcommand{\bra}[1]{\left<#1\right|} 
\newcommand{\ket}[1]{\left|#1\right>} 
\newcommand{\braket}[2]{\left<#1\right.\left|#2\right>} 
 
\newcommand{\eqref}[1]{(\ref{#1})}

\begin{document}
\title{Conditional beam splitting attack on quantum key distribution}
\author{John~Calsamiglia$^1$, Stephen M.~Barnett $^2$, 
 and Norbert~L\"utkenhaus$^3$}
\address{$^1$  Helsinki Institute of Physics, PL 64, 
FIN-00014 Helsingin yliopisto, Finland\\
$^2$ Department of Physics and Applied Physics,
 University of Strathclyde,John Anderson Building,
107 Rottenrow,Glasgow G4 0NG, Scotland.\\
$^3$ MagiQ Technologies Inc.,
275 Seventh Avenue, 26th floor, NY 10001-67 08, USA.}
                
\date{Received \today}
\maketitle
\begin{abstract}

We present a novel attack on quantum key distribution 
based on the idea of \emph{adaptive absorption}\cite{calsam01}. 
The conditional beam splitting attack is shown to be
much more efficient than the conventional beam
spitting attack, achieving a performance similar to the, powerful but 
currently unfeasible, photon number splitting attack. 
The implementation of the conditional beam splitting attack, based solely on 
linear optical elements, is well within reach of current technology.

\end{abstract}
\begin{multicols}{2}

\section{Introduction}

The use of quantum effects  extends our communication capabilities
beyond the solutions offered by classical communication theory. A
prominent example is {\em quantum key distribution} (\QKD) which
allows to expand a small initial secret key shared between two parties
into a larger secret key. This task, which cannot be accomplished
within classical communication theory, enables the two parties to
exchange secret messages via the encryption technique of the {\em one-time
pad} \cite{vernam26a}. 
The idea has been introduced by Wiesner \cite{wiesner83a} and the first
complete protocol for \QKD\ has been given by Bennett and Brassard
\cite{bennett84a}. 

In a quantum optical implementation, the sender (Alice) encodes a
random  bit
value ``0'' and ``1'' in the orthogonal polarization states of a
single photon. She chooses at random either a linear or a circular
polarization basis. The receiver (Bob) uses a polarization bases
chosen at random from these two bases. In the following classical
communication, Alice and Bob identify those signals for which they
used the same basis, and the corresponding bit values form the {\em
sifted key}. Either due to noise or due to eavesdropping, Alice's and
Bob's version of the sifted key differ. As long as the error rate is
below some threshold, they can correct these errors and perform
privacy amplification \cite{bennett95a} to obtain a secure key.
The theoretical security analysis of this scheme has been a subject of
intense research  and only recently a full proof of security
for the whole protocol has been given \cite{mayers98a,shor00a,biham99suba,lo99a}. 

A first implementation of this protocol \cite{bennett92a} demonstrated
the feasibility of this scheme. Since then, many groups improved the
implementations. State of the art schemes can maintain the coherence
of the system over distances as far as 50 km. However, the signals
used in these implementations are not single photon signals. Instead,
weak coherent pulses (\WCP) are used with typical average photon
numbers of $0.1$ or higher.  The signals are described by coherent states in the
chosen polarization mode. This modification of the signals together with the
large loss of the fiber optical implementations over long distances
opens a security loophole \cite{huttner95a}. The restrictions on practical
implementations imposed by the use of \WCP\ signals has been
demonstrated \cite{brassard00a}, giving a limit on the distance over
which \QKD\ can be performed as a function of Bob's detection
efficiency and dark count rate. Nevertheless,  it has been shown that,
despite these restrictions, it is still possible to obtain a secure
key \cite{nl00a,inamorisub00a}

The key process which makes Eve very powerful if Alice and Bob use
\WCP\ signals is the {\em photon number splitting attack} (\PNS)
\cite{brassard00a,nl00a}.  In
this attack Eve performs a {quantum non-demolition measurement} (\QND)
of the {\em total} number of photons of the signal. Whenever she finds
that a signal contains two or more photons, she deterministically
takes one photon out of the signal. The remaining photons of the
signal are then forwarded over a  lossless channel to Bob.
For a \WCP\ with mean photon number $\mu$, sent by
Alice, Eve obtains a single photon in the same polarization state as those
in the signal reaching Bob with probability
\begin{equation}
p_{\PNS}^{\rm{succ}} =1-\rm{e}^{-\mu}-\mu \rm{e}^{-\mu}.
\label{eq:pscPNS}
\end{equation}
Eve now can delay the measurement on that
photon until she learns the polarization basis of each signal,
thereby learning the bit value for each signal.
In order to ensure that Bob does not get too many signals compared to the
installed lossy quantum channel, Eve can actually block some signals
completely, starting with the initial one-photon signals which cannot
be split. We find that this strategy gives the complete key to Eve
once the losses of the installed quantum channel are so high that she
can block all single photon signals. 

 This splitting process used in the \PNS\ attack  is allowed by
quantum mechanics, but the implementation is out of reach of current
technology. Therefore earlier analyses of this situation made use of
the {\em beam splitting attack} (\BS) which has the appeal of
simplicity and feasibility. The basic concept uses the idea that a lossy quantum
channel acts like a combination of a lossless channel and a
beam splitter which accounts for the losses. Eve monitors the second
output arm of the beam splitter and  will gain the
complete knowledge of a bit of the sifted key (via a delayed
measurement) if a multi-photon signal
is split such that Bob and Eve both get at least one photon of the
signal. The central quantities are the probability that Bob receives a
non-vacuum signal
\begin{equation}
P_{\BS}^{\rm{B}}[\neg 0] = 1-\exp \left( - \mu \eta \right)
\label{eq:p0BS}
\end{equation}
where  $\eta$ is the single photon transmission efficiency of
the quantum channel. The probability that Bob and Eve both receive a
signal is
\begin{equation}
p_{\BS}^{\rm{succ}} = \left[1-\exp \left( - \mu \eta \right)\right]
\left[1-\exp \left( - \mu (1-\eta) \right)\right].
\label{eq:pscBS}
\end{equation}
Despite its simplicity and perfect simulation of the lossy channel,
the beam splitter attack is very ineffective when replacing channels
with large losses, i.e. large transmission distances.
In that case, for example, two photon signals are
more likely to see both photons being directed to Eve (and therefore
becoming useless) rather than being split. 

In \cite{calsam01} the authors present the idea of \emph{adaptive absorption}.
This consists of sending a photonic signal through a linear absorber 
in which absorption events can be continuously monitored in such a 
way that as soon as a single photon is absorbed, a feed-forward 
mechanism decouples the signal from the absorbing medium. 
With this simple procedure 
it is possible to extract precisely one photon from a field-mode prepared in 
any state (other than the vacuum, of course). 
In this paper we show how the idea of adaptive absorption leads 
to the \emph{conditional beam splitting attack} (\CBS) 
on weak coherent pulse \QKD. 
By using only linear optical elements, 
the \CBS\ reduces the number of 
events where more than one photon is split off. This allows an 
eavesdropping efficiency that overwhelms the conventional beam 
splitting attack and can be as large as the ones given by the \PNS.

The paper will be organized as follows. In Sec.~\ref{sec:CBS} we 
describe the \CBS\ attack and introduce the quantum jump method, 
which in turn will be used to calculate the  state of the signal
during the various stages of the attack. The results will allow us
to compare the performances of the \CBS\ and the conventional \BS\ 
attack. For sake of simplicity this will be done in the scenario in 
which the eavesdropper is able delay her measurement until the 
encoding basis is announced by Alice \cite{note1}. In Sec.~\ref{sec:DCBS} we 
consider the more realistic situation in which Eve does not have the 
technological skills to store photons, and introduce
a variation of the \CBS\ where Eve tries to split two single photons 
from the signal before forwarding it to Bob. 
In Sec.~\ref{sec:phstat} we study the photon statistics in Bob's 
detectors and see that in principle Alice and Bob could use this
information to disclose Eve's attack.
The possibility of improving the \CBS\ attack by 
using mixed strategies will be investigated 
in Sec.~\ref{sec:mixstrat}. 
In  Sec.~\ref{sec:CBSF} we rederive some basic
results for finite beam splitters  and compare them
with the ones obtained using the quantum jump method.
Sec.~\ref{sec:concl} concludes the paper with a brief summary.

\section{Conditional beam splitting attack} \label{sec:CBS}

As mentioned in the introduction the hope is to take advantage of 
the fact that the signal bits are implemented through polarized coherent states
with very low mean photon number (instead of single photons). 
To do that Eve will weakly couple her modes   to the signal modes 
and try to extract one excitation from the signal sent 
by Alice. As soon as she gets one photon into her modes,
Eve will allow any remaining signal photons  
to reach Bob through an ideal channel. Otherwise she will keep on trying for 
a longer time.
 If she does not succeed after a maximum coupling time $\tau$,
Eve will directly send the signal to Bob through the ideal channel.

After Alice announces publicly the encoding basis used to send each of the 
bits, Eve can measure her photon to learn the bit value of transmitted signal 
\cite{note1}.
Only in the cases where the multiphoton signal is split in such a way that both 
Eve and Bob receive a non-vacuum signal, will the bit value learned by Eve 
form part of the sifted key shared between Alice an Bob. We will 
therefore refer to the probability of this event as the probability 
of success of the \CBS\ ($p^{\rm{succ}}_{\CBS}$). 
Since this attack
does not produce any qubit errors, we will take the probability of 
success as a figure of merit for the attack.
On the other hand, in order to remain unnoticed,
Eve's attack has to be such that the 
number of non-vacuum signals that arrive to Bob agrees with 
what he expects from the lossy channel. Hence, the probability 
$P^{\rm{B}}_{\CBS}[\neg 0]$ that Bob receives a non-vacuum
 signal fixes a bound on the 
eavesdropping attack. The probabilities $p^{\rm{succ}}_{\CBS}$ 
and $P^{\rm{B}}_{\CBS}[\neg 0]$ will be the central quantities when 
evaluating the attack.

\begin{figure}
    \centerline{\psfig{figure=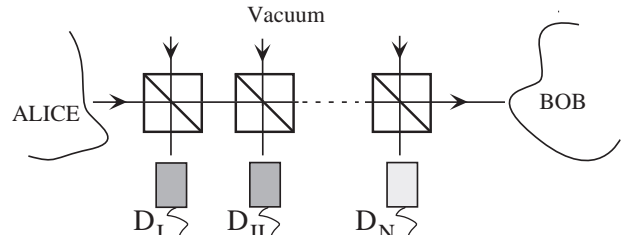,width=3.2in}}
    \caption{Possible experimental realization of the conditional beam 
    splitting attack.}
    \label{fig:impl}
\end{figure}

In Fig.\ref{fig:impl} a possible implementation of the \CBS\ is 
shown. The initial state sent by Alice 
occupies only two photonic modes ($a$ and $b$) 
corresponding for example to the two polarization degrees of freedom 
of a traveling mode. 
Conditional beam splitting 
consists of sending the input state to a 
polarization independent weak beam splitter. A measurement to determine if
there are any photons is then done in the weakly coupled output arm 
(modes $a_{e}$ and $b_{e}$).
If no photon is detected the signal is sent through an identical 
beam splitter again. Otherwise the signal is transmitted through a 
perfect channel without any further processing. 

To investigate this procedure we will take the limit of infinitesimally
weak beam splitting which will allow us to take results 
from quantum jump methods
\cite{molmer93,plenio98,carmich93} as we did for the study of adaptive 
absorption \cite{calsam01}. In Sec.~\ref{sec:CBSF} we will give 
some numerical results for finitely weak beam splitters.
Quantum jump methods are wave function (as opposed to 
density matrix) approaches  to study the
evolution of small systems coupled to a large reservoir. 
In addition to being a very powerful method 
it has a nice physical interpretation: The stochastic evolution of the 
wave function corresponds to the stochastic read outs of the 
continuously monitored reservoir.

We preface our analysis with a brief review of
 the quantum jump 
approach to open systems.
Suppose that initially the system is in the state $\ket{\phi(0)}$.
If no jump occurs the evolution of the system is described by the 
effective Hamiltonian,
 $H_{\rm{eff}}=H_{S}-\frac{i}{2}\sum_{m}J^{\dagger}_{m}J_{m}$, 
\begin{equation}
    \ket{\phi^{0}(t)}=\mbox{e}^{-i H_{\rm{eff}} t}\ket{\phi(0)}\mbox{.}
    \label{eq:nojump}
\end{equation}
Here $H_{S}$ is the Hamiltonian of the isolated system and the 
$J_{m}$'s are the jump operators which account for the coupling to 
the reservoir. Since the effective Hamiltonian is not hermitian, this state
$\ket{\phi^{0}(t)}$ is not normalized. 
The square of its norm gives the probability of having no jump 
after a time $t$,
\begin{equation}
    p_{0}(t)=\bra{\phi(0)}\mbox{e}^{i
    H_{\rm{eff}}^{\dagger} t}{e}^{-iH_{\rm{eff}} t}\ket{\phi(0)}
    \label{eq:prbnojump}
\end{equation}
If a jump to a mode $m$ occurs in a time between $t$ and $t+\delta t$ 
then the
system will be in the state
\begin{equation}
    \ket{\phi^{m}(t+\delta t)}=\sqrt{d t}J_{m}\ket{\phi(t)}
    \label{eq:jump}
\end{equation}
immediately after this period. 
And again the probability of this event is given by the square of the 
norm of the conditional state, 
\begin{equation}
    \delta p_{m}(t)=\bra{\phi(t)}J_{m}^{\dagger}J_{m}\ket{\phi(t)}d t
    \label{eq:deltaprb}
\end{equation}
Using these probabilities one can follow the history of the system's wave 
function when the reservoir is continuously monitored to detect what 
kind of jump occurred or if any jump occurred at all. It can be shown 
that by averaging over different histories 
the state of the system at a given time $t$, one 
recovers the master equation

\begin{eqnarray}
 \frac{d \rho}{dt}&=&-i[H_{S},\rho]-\frac{1}{2}\sum_{m}(J^{\dagger}_{m}J_{m}\rho
    +\rho J^{\dagger}_{m}J_{m})\nonumber\\
    & &+\sum_{m}J_{m}\rho J_{m}^{\dagger}
    \label{eq:mastereq}
\end{eqnarray}
for the system density operator $\rho$.
The first term in this equation describes the standard unitary 
evolution of the system.
The last two terms  describe the relaxation process due 
to the coupling of the system to the reservoir.
In order for this description to be true, the coupling and the reservoir have to 
be such that the jump probability  $\delta p_{m}(t)$  is very small 
and does not depend on the previous history of the system. 

In this paper we use the quantum jump method to study the
evolution of the system formed by the \emph{signal modes} ($a$ and 
$b$), and Eve's modes ($a_{e}$ and $b_{e}$).
The action of the infinitesimally weak beam splitting together with a 
measuring device that checks if the number of photons in Eve's modes 
has increased will play the role of reservoir being continuously monitored.
An increase in the photon number in Eve's modes is then 
represented by the jump operator
\begin{equation}
    J=\epsilon(a_{e}^{\dagger}a+b_{e}^{\dagger}b)\mbox{,}
    \label{eq:jumpop}
\end{equation}
where $\epsilon\ll 1$ is the reflection coefficient of the weak 
beamsplitting. After this jump the photons in Eve's modes and the signal 
photons will be in a shared wave function. If no jump occurs Eve's 
modes will remain in the vacuum state.

The initial state is given by
\begin{equation}
    \ket{\phi(0)}=\ket{\alpha; \beta}\ket{0;0}\mbox{,}
    \label{eq:init}
\end{equation}
where the first two modes are the signal modes initially in a 
coherent state and with a definite polarization, 
and the last modes are Eve's modes. The mean photon 
number of the signal sent by Alice is $\mu=|\alpha |^2+|\beta |^2$.
If Eve does not intervene then the signal that Bob will receive after 
going through the lossy channel is
\begin{equation}
    \ket{\phi^B_{\eta}}=\ket{\sqrt{\eta} \alpha; \sqrt{\eta} \beta}\ket{0;0},
    \label{eq:nointerv}
\end{equation}
where $\eta$ is the transmissivity of the channel. This means that the 
probability that Bob gets a non-vacuum signal is
\begin{equation}
    P^{\rm{B}}_{\eta}[\neg 0]=1-P^B_{\eta}[0]=1-\exp (-\mu\eta),
    \label{eq:p0B}
\end{equation}
where $P^{\rm{B}}_{\eta}[0]$ is the probability of Bob getting the vacuum signal.

We now proceed to calculate what happens when Eve tries to 
eavesdrop on the signal using the conditional beam splitting attack (\CBS).
At time $t_{o}=0$ Eve starts a conditional beam splitting attack on the 
signal sent by Alice. The conditional state of the system after a time $t$ 
when no photon has been detected in Eve's modes is
\begin{eqnarray}
\ket{\phi^{0}(t)} & = & \mbox{e}^{-t\frac{1}{2}J^{\dagger}J}\ket{\phi(0)}
     \nonumber  \\
     &=&\mbox{e}^{-\frac{t \epsilon^{2}}{2} 
     (n_{a}+n_{b})}\ket{\alpha; \beta}\ket{0;0}\nonumber  \\
     & = & \mbox{e}^{\frac{1}{2}(\gamma_{t}^{2}-1)\mu}
     \ket{\gamma_{t}\alpha; \gamma_{t}\beta}\ket{0;0}\mbox{,}
    \label{eq:nojump2}
\end{eqnarray}
where we have defined $\gamma_{t}=\exp (-\frac{t\epsilon^{2}}{2})$ and 
used the normal ordered form of the exponential of the number operator 
\cite{barnett97}. The 
squared norm of this state is the probability of detecting no photon 
in Eve's mode after a time $t$,
\begin{equation}
    p_{0}(t)=\braket{\phi^{0}(t)}{\phi^{0}(t)}=
    \mbox{e}^{(\gamma_{t}^{2}-1)\mu}\mbox{.}
    \label{eq:prbnophoton}
\end{equation}

If the first jump occurs in the time interval $[t,t+d t]$ then the 
conditional state of the system is
\begin{eqnarray}
 & & \ket{\phi^{1}(t+d t )}  = J 
 {\rm e}^{-t\frac{1}{2}J^{\dagger}J}\ket{\phi(0)}
   \nonumber  \\
     & &=  {\rm e}^{\frac{1}{2}(\gamma_{t}^{2}-1)\mu} J
     \ket{\gamma_{t_{1}}\alpha; \gamma_{t}\beta}\ket{0;0} \nonumber   \\
      & & =  {\rm e}^{\frac{1}{2}(\gamma_{t}^{2}-1)\mu} \epsilon \gamma_{t_{1}}
     \ket{\gamma_{t}\alpha; \gamma_{t_{1}}\beta}(\alpha \ket{1;0}+
      \beta \ket{0;1})\mbox{.}
    \label{eq:jump2}
\end{eqnarray}
The probability density of detecting a photon in Eve's modes at the time interval 
$[t,t+\delta t]$ is therefore given by 
\begin{equation}
    p_{1}(t)=\braket{\phi^{1}(t)}{\phi^{1}(t)}=\epsilon^2\gamma_{t_{1}}^2\mu 
    \mbox{e}^{(\gamma_{t_{1}}^{2}-1)\mu}\mbox{.}
    \label{eq:prbjump}
\end{equation}

As already mentioned,
the first thing that Alice and Bob will check to detect the presence of Eve,
is that Bob receives a fraction of non-vacuum signals
consistent with the lossy channel, given in ~\eqref{eq:p0B}. That is,
\begin{equation}
     P^{{\rm B}}_{\CBS}[\neg 0]  =  P^{{\rm B}}_{\eta}[\neg 0] 
   \;  \mbox{  or  } \; P^{{\rm B}}_{\CBS}[0]  =  P^{{\rm B}}_{\eta}[0]\mbox{.}
    \label{eq:p0cond}  
\end{equation}
The total probability that after the \CBS\ attack Bob 
gets a vacuum signal is 
\begin{eqnarray}
    P^{{\rm B}}_{\CBS}[0] & = & 
    p_{0}(\tau) |\!\braket{0;0}{\gamma_{\tau}\alpha;\gamma_{\tau}\beta}|^2 +
    \nonumber\\ 
   & &+ \int_{0}^{\tau}p_{1}(t)|\!\braket{0;0}{\gamma_{t}\alpha;\gamma_{t}\beta}|^2 
  dt   \nonumber\\
  &=& {\rm e}^{-\mu}(1+\mu(1-\gamma_{\tau}^2)),
    \label{eq:p0g}
\end{eqnarray}
where we have made use of Eqs. \eqref{eq:prbnophoton} and \eqref{eq:prbjump}.
With the results of Eqs.~\eqref{eq:p0B} and \eqref{eq:p0g} 
we find the required value of the coupling time $\tau$ so that 
condition expressed in Eq.~\eqref{eq:p0cond} is fulfilled, 
\begin{equation}
 \gamma_{\tau}^2={\rm e}^{-\epsilon^2 
    \tau}=\frac{1}{\mu}(1+\mu-{\rm e}^{\mu(1-\eta)})\mbox{. }
    \label{eq:g2t}
\end{equation}
Notice that $\gamma_{\tau}^2<\eta$.

In order to quantify the  performance of the \CBS\ we will now 
calculate the probability of a successful 
splitting $p^{{\rm succ}}_{\CBS}$. This is the probability that Eve manages 
to extract one photon from the signal and still leaves at least one 
photon for Bob. Splittings that leave the transmitted signal in the 
vacuum state are not useful to Eve since these bits will not 
contribute to the sifted key. The success probability for the \CBS\ 
attack is given by
\begin{eqnarray}
   p^{{\rm succ}}_{\CBS} & = & 1-(P^{{\rm B}}_{\CBS}[0]+ 
   P^E_{\CBS}[0]-P^{{\rm EB}}[0,0]) 
    \label{eq:psucc}  \\
     & = & 1-{\rm e}^{-\mu}(1+\mu(1-\gamma_{\tau}^2))-
     {\rm e}^{(\gamma_{\tau}^2-1)\mu}+{\rm e}^{-\mu}
    \nonumber  \\
     & = & 1-\mu{\rm e}^{-\mu}(1-\gamma_{\tau}^2)-
     {\rm e}^{-(1-\gamma_{\tau}^2)\mu},
    \label{eq:psucccbs}
\end{eqnarray}
where  $P^E_{\CBS}[0]=p_{0}(\tau)$ is the probability of having 
no photon in Eve's modes (i.e. no jump) after the attack and
$P^{{\rm EB}}[0,0]={\rm e}^{-\mu}$ is the probability that there 
are photons neither in Eve nor in Bob's modes.

By inverting  Eq.~\eqref{eq:g2t} we find the transmissivity $\eta$
`mimicked' (in the sense of Eq.~\eqref{eq:p0cond}) by the \CBS\ attack,
\begin{equation}
  \eta_{\CBS}=  1-\frac{1}{\mu}\ln(1+\mu(1-\gamma_{\tau}^2)).
  \label{eq:etaCBS}
\end{equation}
Since $\eta_{\CBS}$ is an increasing function of $\gamma_{\tau}^2$ and we know 
that this achieves its minimum 
for $\tau\rightarrow\infty$ ($\gamma_{\tau}^2\rightarrow 0$),
we find that, given a mean photon 
number $\mu$, the attack just described cannot mimic arbitrarily
large channel losses. The  minimum transmissivity is then given by
\begin{equation}
  \eta_{\CBS}^{{\rm min}}=  1-\frac{1}{\mu}\ln(1+\mu) \approx 
  \frac{1}{2}\mu-\frac{1}{3}\mu^2+O(\mu^3)\mbox{.}
  \label{eq:etaCBSmin}
\end{equation}
This makes sense since for $\tau\rightarrow\infty$ all non-vacuum
signals will leak one photon to Eve while the remaining photons always
reach Bob. It is clear, therefore, that the removal of only one photon 
cannot account for arbitrarily high channel losses. 
In order to meet Bob's expectations even when $\eta\leq\eta_{\CBS}^{{\rm min}}$,
Eve can apply the 
protocol corresponding to $\eta_{\CBS}^{{\rm min}}$ and then block the 
outgoing signals with a probability
\begin{equation}
     p^{{\rm block}}_{\CBS}=\frac{1-{\rm e}^{\mu\eta}}
     {1-{\rm e}^{\mu\eta_{\CBS}^{{\rm min}}}}\mbox{.}
    \label{eq:pblock}
\end{equation}
Note that if the channel loss is equal to or larger than 
 $1-\eta_{\CBS}^{{\rm min}}$, then 
the probability of success is equal to the 
probability of having more than one photon in a signal pulse
\begin{equation}
    p^{{\rm succ}}_{\CBS}(\eta_{\CBS}^{{\rm min}})=
    1-{\rm e}^{\mu}-\mu{\rm e}^{\mu}\mbox{.}
    \label{eq:psuccmin}
\end{equation}
This means that for these high channel losses (i.e. $\eta\leq
\eta_{\CBS}^{{\rm min}}$) Eve can extract one excitation from the signal
without modifying Bob's expected number of non-vacuum signals. 
All Bob's non-vacuum contributions effectively come from the multiphoton part of 
the signal pulses, and Eve will possess one photon from each of those 
signals. Therefore she will obtain the 
\emph{full} sifted key shared by Alice and Bob after the public 
announcement of the bases. This could never have happen if Eve had chosen
the \BS\ attack. This is a very important feature since  Eve's 
knowledge of the full key does not leave any room for Alice and Bob 
to perform  privacy amplification to obtain a secure key.

\begin{figure}
    \centerline{\psfig{figure=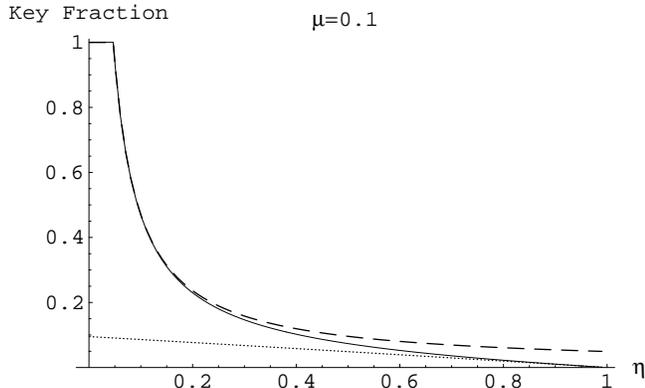,width=3.4in}}
    \caption{Fraction of the key known by Eve in \CBS\ (solid 
    curve), \BS\ (dotted) and \PNS\ (dashed) attacks for $\mu=0.1$}
    \protect\label{fig:frgfrbs}
\end{figure}

Taking into account the blocking, in the regime of high losses, the success 
probability calculated in ~\eqref{eq:psucccbs} 
takes the following form
\begin{eqnarray}
   p^{{\rm succ}}_{\CBS} =
    \left\{
    \begin{array}{cc}
      1-\mu{\rm e}^{-\mu}(1-\gamma_{\tau}^2)-{\rm e}^{-(1-\gamma_{\tau}^2)\mu}
      & \mbox{: } \eta>\eta_{\CBS}^{{\rm min}}  \\
      \\
       p^{{\rm block}}_{\CBS}\lim\limits_{\gamma_{\tau}\to 0} p^{{\rm succ}}_{\CBS}
	 =1-{\rm e}^{-\mu \eta} 
      &\mbox{: } \eta\leq\eta_{\CBS}^{{\rm min}} 
    \end{array}
    \right.\mbox{.}
    \label{eq:pscgT}
\end{eqnarray}
In order to evaluate the performance of an attack it is more 
convenient to normalize the probability of success with the 
probability that Bob gets a non-vacuum signal (potential sifted key 
bit) to obtain the \emph{key fraction} known by Eve,
\begin{equation}
    f_{\CBS}=\frac{p^{{\rm succ}}_{\CBS}}{1-P^{{\rm B}}_{\CBS}[0]}\mbox{.}
    \label{eq:fg}
\end{equation}
As discussed previously, for $\eta\leq\eta_{\CBS}^{{\rm min}}$ Eve can acquire 
the whole key, so in this regime $f_{\CBS}=1$. For other channel loss values 
the key fraction never reaches unity. 
Similarly one can define the same quantity for the \BS\ attack 
obtaining
\begin{equation}
    f_{\BS}=\frac{p^{{\rm succ}}_{\BS}}{1-P^{{\rm B}}_{\BS}[0]}=
    1-{\rm e}^{-(1-\eta)\mu} , 
    \label{eq:fbs}
\end{equation}
where we have made use of Eq.~\eqref{eq:p0BS} and ~\eqref{eq:pscBS}.
It is easy to prove that the fraction of key known by Eve is always 
bigger for the \CBS\ than for the \BS\ attack.
In Fig.~\ref{fig:frgfrbs}  we can see these fractions as a function of the 
channel transmissivity for a typical value of the mean photon number 
used in current experiments (see also Fig.~\ref{fig:frqa}). 

\begin{figure}[htbp]
    \centerline{\psfig{figure=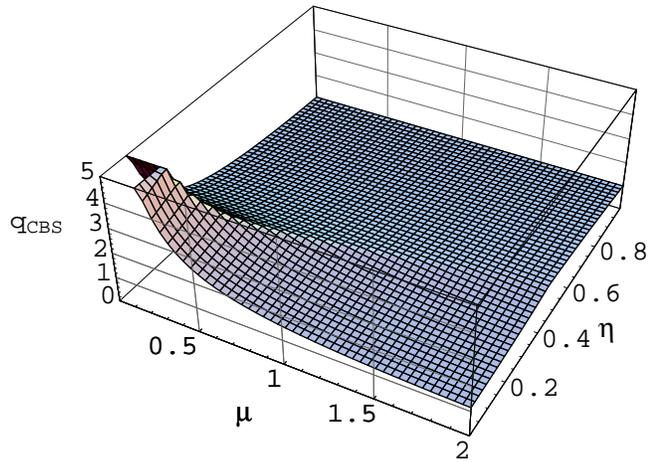,width=3.4in}}
    \caption{Performance quotient $q_{\CBS}$ for the \CBS. Since it is 
    not bounded from above we have only plotted values smaller than 
    five.} 
    \label{fig:qcbs1}
\end{figure}

In order to compare quantitatively the \CBS\ with the \BS\ 
we define the performance quotient $q_{\CBS}=\frac{f_{\CBS}}{f_{\BS}}$
which is plotted in Fig.~\ref{fig:qcbs1} as a function of the mean 
photon number of the signal pulses and the channel transmissivity.
\noindent We see that for small $\mu$ the \CBS\  can be substantially 
better than the \BS. For example, for the experimentally reasonable 
values $\mu=0.1$ and $\eta=0.1$ the \CBS\ provides Eve with a 
fraction of the key $q_{\CBS}=5.4$ times bigger than the \BS\ attack.
For a fixed mean photon number $\mu$, the optimum advantage 
$q^{{\rm max}}_{\CBS}=1+\mu^{-1}$ 
is achieved when the channel losses are $\eta=\eta_{\CBS}^{{\rm min}}$.

\section{CBS without storage} \label{sec:DCBS}
There is an important fact that makes the \CBS, as described above, 
rather more technically demanding than the \BS. 
In order to follow the \CBS\ protocol 
described in the previous section Eve has to be able to perform two 
experimentally non-trivial tasks \cite{note2}.
\emph{a)} Firstly she has to detect the presence of photons in her modes $a_{e}$ and $b_{e}$ 
without altering their polarization. This is necessary because Eve has 
to be able to carry out the conditional dynamics, i.e. stop the 
splitting as soon as she gets the desired photon. This operation is 
not  as technologically demanding as the \PNS\ since it only needs to 
discriminate the non-vacuum states from the vacuum. However, it is  
out of reach of the immediately available technology.
\emph{b)} The second task Eve must be able to realize is to 
store the extracted photon until the encoding basis is publicly announced by 
Alice. 

In the conventional beam splitting attack Eve is not required 
to carry out task \emph{a)}. On the other hand, we notice that  Alice 
can always delay the public announcement, making it harder for
Eve to store coherently the signal and thereby, effectively  
constraining Eve to attacks that do not rely on 
the storage of the extracted signal.
This of course applies to all the error-free attacks that we have 
seen in this paper, that is the \PNS, \BS\ and \CBS. 
Inasmuch as Eve is forced to carry out the signal measurement before knowing the 
encoding basis,  in the \CBS\ attack she does not have to rely anymore 
on her ability to detect photons without disturbing their 
polarization  (task \emph{a)}).
Eve can realize the conditional dynamics by directly 
measuring the extracted photons with photodetectors.
Acknowledging the impracticability of indefinite qubit storing, thus, 
puts on equal footing the \BS\ and \CBS\ as far as technological difficulty 
is concerned, and still leaves the \PNS\ as unfeasible.

As in the previous section, we can now calculate the performance
of the \CBS\ in this new no-storage or direct measurement scenario. 
In this situation
the \CBS\ is bound to fail in half of the cases. Eve only succeeds 
when she measures the extracted photon in the right basis. The 
probability of success and key fraction in the \emph{directly measured 
conditional beam splitting} (\DCBS) will accordantly be reduced by a 
factor $\frac{1}{2}$, i.e.
$p^{{\rm succ}}_{\DCBS}= \frac{1}{2}p^{{\rm succ}}_{\CBS}$ and
$f_{\DCBS}= \frac{1}{2}f_{\CBS}$. 
Clearly, since the number of extracted photons is the same as 
in the scenario with the possibility of storage, the number of non-vacuum signal 
that arrive at Bob's site remains the same. 

The success probability for the directly measured beam splitting attack (\DBS )
 can be calculated taking into account that Eve's 
attack is unsuccessful only in the case where all split photons from a signal 
are measured \cite{note3}
in the wrong basis,
\begin{eqnarray}
    p^{{\rm succ}}_{\DBS} &=&(1-e^{-\eta 
    \mu})\left(1-\sum_{n=0}^{\infty}\frac{1}{2^{n}}\frac{\mu^{n}(1-\eta)^{n}}{n!}
    e^{-\mu(1-\eta)}\right) \nonumber \\ 
    & = & (1-e^{-\eta \mu})(1- e^{-\frac{\mu}{2}(1-\eta)})\mbox{.}
    \label{eq:psDCBS}
\end{eqnarray}
The fraction of the key known by Eve in this attack is therefore,
\begin{equation}
    f_{\DBS}=1- e^{-\frac{\mu}{2}(1-\eta)}\mbox{.}
    \label{eq:frDBS}
\end{equation}

\begin{figure}
    \centerline{\psfig{figure=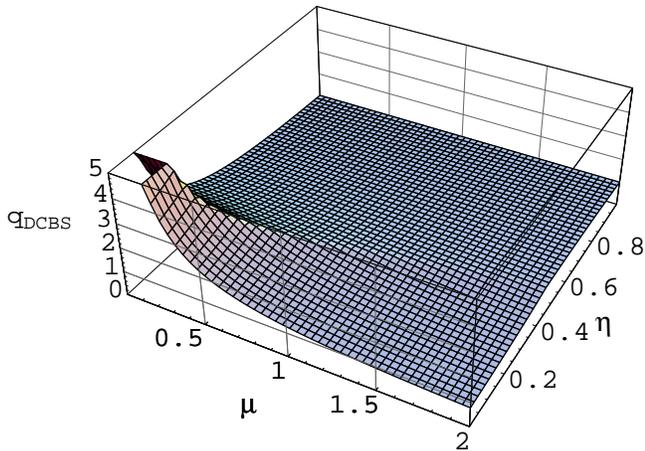,width=3.4in}}
    \caption{Performance quotient $q_{\DCBS}$ for the $\DCBS$ . Since it is 
    not bounded from above we have only plotted values smaller than 
    five.}
    \protect\label{fig:qDCBS1}
\end{figure}

\begin{figure}[!hbp]
    \centerline{\psfig{figure=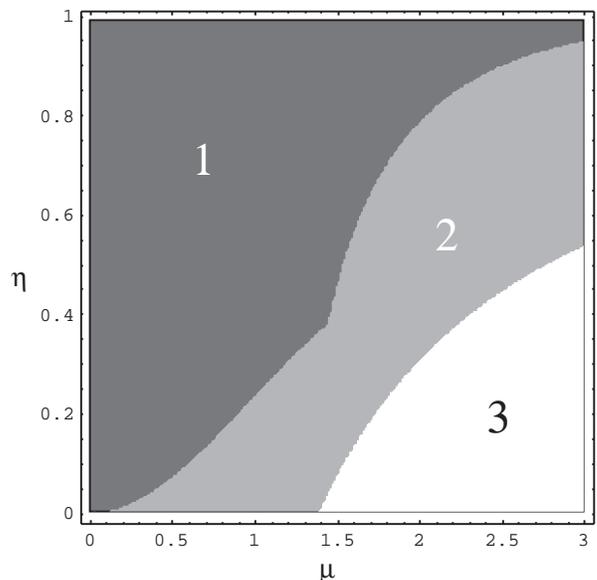,width=3.1in}}
    \caption{Diagram of dominance of the different attacks under 
    no-storing conditions: 1.$ f_{\DCBS}\geq f_{\ACBS}\geq f_{\DBS}$, 
    2.$ f_{\ACBS}\geq f_{\DCBS}\geq f_{\DBS}$ and 3.$ f_{\ACBS}\geq 
    f_{\DBS}\geq f_{\DCBS}$  }
    \protect\label{fig:reg}
\end{figure}

\begin{figure}[!hbp]
    \centerline{\psfig{figure=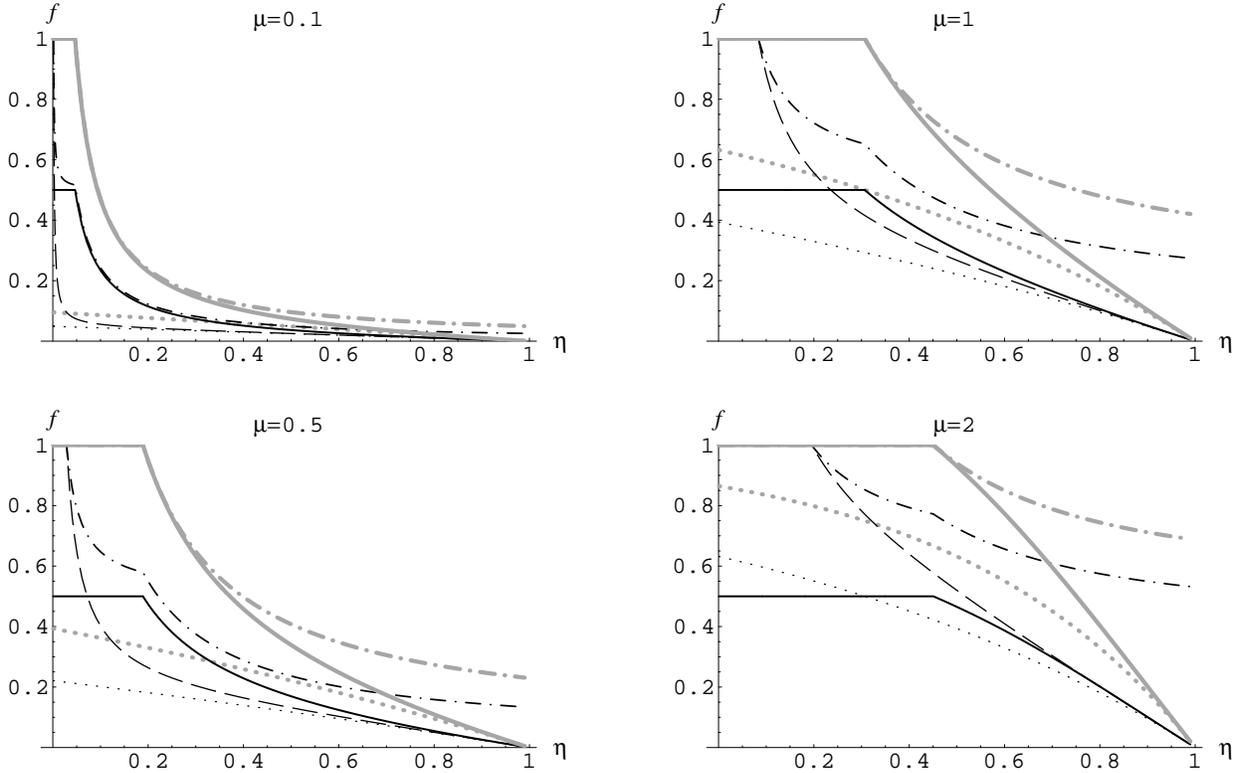,width=7in}}
    \caption{Key fraction for different scenarios 
	(Black: no-storage allowed, Gray: storage allowed) 
    and attacks (Solid: \CBS, Dashed: \ACBS, Dotted: \BS, Dot-dash: 
    \PNS).  }
    \protect\label{fig:frqa}
\end{figure}

Since for small $\mu$ the most important contribution in the \BS\ 
comes from the two photon signals, as in \CBS, the success probability 
of both attacks is reduced approximately by the same factor $\frac{1}{2}$. 
But this factor is always a bit larger for \DBS\ since in the cases where 
more than one photon per signal are split, Eve has a bigger chance 
to measure in the right basis. The performance quotient is now defined 
relatively to the \DBS, $q_{\DCBS}=
\frac{f_{\DCBS}}{f_{\DBS}}$ and it is plotted in Fig.~\ref{fig:qDCBS1}
 for relevant values of $\mu$ and $\eta$. We see that 
for large mean photon numbers ($\mu>\ln 4\approx 1.4$) the \DBS\ can 
actually perform slightly better than the $\DCBS$ for some range of channel 
losses (see also Figs.~\ref{fig:frqa} and \ref{fig:reg}).
\noindent Despite this last remark it looks like 
the \CBS\ maintains its efficiency over the \BS\ 
under the no-storage constraint (compare 
Fig.~\ref{fig:qDCBS1} with Fig.~\ref{fig:qcbs1}), but in 
fact under this scenario the $\DCBS$ has lost the threatening 
feature of being able to extract the full key ($f_{\CBS}=1$) for 
high channel losses (see Fig.~\ref{fig:frqa}).

In the remaining of this section we will propose a variation of the 
\CBS\ which allows Eve to extract the full key even in the no-storage 
scenario, and to perform better than \DBS\ for any mean photon 
number. 
The idea of the adapted conditional beam 
splitting attack (\ACBS) is to extract two photons, one by one, from the 
signal and measure each of them in a different polarization basis. In 
order to do that, Eve just has to follow the same protocol as in the 
single photon \CBS\ but instead of stopping the beam splitting as 
soon as she detects one photon, she has to continue the splitting until 
a second photon is extracted. As previously, in order to match the 
expected number of photons in Bob's side, the splitting procedure can
run for a maximum coupling time $\tilde{\tau}$ after which the signal must be 
transmitted to Bob through a lossless channel.
Obviously, this attack will only be advantegeous in the no-storage scenario 
since otherwise, when the key can be extracted from the first photon,
it is of no use to extract a second photon.

The total probability that  Bob 
gets a vacuum signal after the  \ACBS\ is 
\begin{eqnarray}
   && P^{{\rm B}}_{\ACBS}[0]  =  
    p_{0}(\tilde{\tau}) p_{c}^0(\mu)
   +\int_{0}^{\tilde{\tau}}p_{1}(t_{1}) \nonumber\\
  & & \left[ \int_{t_{1}}^{\tilde{\tau}} p_{11}(t_{2}|t_{1})
   p_{c}^0(\gamma_{t_{2}}^2\mu) dt_{2}  
  + p_{10}(\tilde{\tau}|t_{1}) 
  p_{c}^0(\gamma_{\tilde{\tau}}^2\mu)\right] dt_{1} 
   \nonumber\\ 
   & &= {\rm e}^{-\mu}(1+\mu(1-\gamma_{\tilde{\tau}}^2)+
   \frac{\mu^2}{2}(1-\gamma_{\tilde{\tau}}^2)^2)\mbox{,}
    \label{eq:p02g}
\end{eqnarray}
where $p_{c}^n(\mu)$ is the probability of having $n$ photons in a 
coherent state with mean photon number $\mu$, 
$p_{11}(t_{2}|t_{1})=\epsilon ^2\gamma_{t_{2}}^2\mu 
{\rm e}^{\mu(\gamma_{t_{2}}^2-\gamma_{t_{1}}^2)}$ is 
the probability of having a jump at $t_{2}$ conditional to a previous 
jump at $t_{1}$, and $p_{10}(\tilde{\tau}|t_{1})=
{\rm e}^{\mu(\gamma_{\tau}^2-\gamma_{t_{1}}^2)}$ is the probability 
of having no jump in the time interval $[t_{1},\tilde{\tau}]$ 
conditional to a jump at $t_{1}$. These probabilities can be 
calculated following the general results in the beginning of  
Sec.~\ref{sec:CBS}.
The previous result is equal to the corresponding probability in \CBS\ 
(Eq.\ref{eq:p0g}) plus 
a second order in $\mu$ term which represents the removal of two 
photons.

The coupling time $\tilde{\tau}$ can be now fixed so that 
Bob's probability of detecting at least one photon agrees with the 
result he would expect from the lossy channel, 
\begin{eqnarray}
    P^{{\rm B}}_{\ACBS}[\neg 0] & = & P^{{\rm B}}_{\eta}[\neg 0] \longrightarrow P^{{\rm B}}_{\ACBS}[0]  =  P^{{\rm B}}_{\eta}[0]
    \label{eq:p0condb}  \\
    \longrightarrow \gamma_{\tilde{\tau}}^2 & =& {\rm e}^{-\epsilon^2 
    \tilde{\tau}}=1+\frac{1}{\mu}
    (1-\sqrt{2{\rm e}^{\mu(1-\eta)}-1})\mbox{.}
    \label{eq:g2tb}
\end{eqnarray}
As expected we see that the maximum coupling time will be smaller for 
the \ACBS\ than the \CBS\, $ \gamma_{\tau}^2<\gamma_{\tilde{\tau}}^2 <\eta$.

To calculate the probability of success we have to count the events
in which Eve extracts a signal while leaving some non-vacuum 
contribution to Bob. We also have to take into account that Eve only 
gets the bit value with certainty when she manages to extract two 
single photons, otherwise she will only get it in half of the cases.
The success probability for the \ACBS\ attack is then given by
\begin{eqnarray}
  & & p^{{\rm succ}}_{\ACBS}  =  \int_{0}^{\tilde{\tau}}p_{1}(t_{1})
   \left[\int_{t_{1}}^{\tilde{\tau}} p_{11}(t_{2}|t_{1})
   p_{c}^{\neg 0}(\gamma_{t_{2}}^2\mu) dt_{2} + \right.\nonumber\\
  & &\left. +\frac{1}{2} p_{10}(\tilde{\tau}|t_{1}) 
  p_{c}^{\neg 0}(\gamma_{\tilde{\tau}}^2\mu)\right] dt_{1} 
  = 1-{\rm e}^{-\mu}\left[{\rm e}^{\gamma_{\tilde{\tau}}^2\mu}+\right.
  \nonumber\\ 
   &  & \left. +\frac{\mu}{2}
   (1-\gamma_{\tilde{\tau}}^2) (1+{\rm e}^{\gamma_{\tilde{\tau}}^2\mu})+
   \frac{\mu^2}{2}(1-\gamma_{\tilde{\tau}}^2)^2 \right] \mbox{,}
    \label{eq:psucACBS}
\end{eqnarray}
where $p_{c}^{\neg 0}(\mu)$ is the probability of having at least one
photon in a coherent state with mean photon number $\mu$.

By inverting  Eq.~\eqref{eq:g2tb} we find that the transmissivity 
`mimicked' (in the sense of Eq.~\eqref{eq:p0condb}) by the \ACBS\ attack 
is
\begin{equation}
  \eta_{\ACBS}=  
  1-\ln(1+\mu(1-\gamma_{\tilde{\tau}}^2)+
  \frac{\mu^2}{2}(1-\gamma_{\tilde{\tau}}^2)^2).
  \label{eq:etaACBS}
\end{equation}
Since this is an increasing function of $\gamma_{\tilde{\tau}}^2$ we 
find the minimum transmissivity that can be mimicked by the \ACBS\ 
without extra blocking (for $\tilde{\tau}\rightarrow\infty$) is
\begin{equation}
  \eta_{\ACBS}^{{\rm min}}=  1-\frac{1}{\mu}\ln(1+\mu+\frac{\mu^2}{2}) \approx 
  \frac{1}{2}\mu^2+O(\mu^3) .
  \label{eq:etaACBSmin}
\end{equation}
 
In this limit of very high losses 
Eve can extract two excitations from all signals and 
still meet Bob's expectations.
Therefore, by measuring each photon in a different basis,
she will be able to acquire the full key even in the 
no-storage scenario (
$f_{\ACBS}=\frac{p^{{\rm succ}}_{\ACBS}}{P^{{\rm B}}_{\ACBS}[\neg 0]}=1$). 
If the losses are still higher  ($\eta\leq\eta_{\ACBS}^{{\rm min}}$)
Eve has to block some signals with probability
\begin{equation}
    p^{{\rm block}}_{\ACBS}=
    \frac{1-{\rm e}^{-\mu\eta}}{1-{\rm e}^{-\mu\eta_{\ACBS}^{{\rm min}}}}.
    \label{eq:pblockACBS}
\end{equation}

\begin{figure}[!hbp]
    \centerline{\psfig{figure=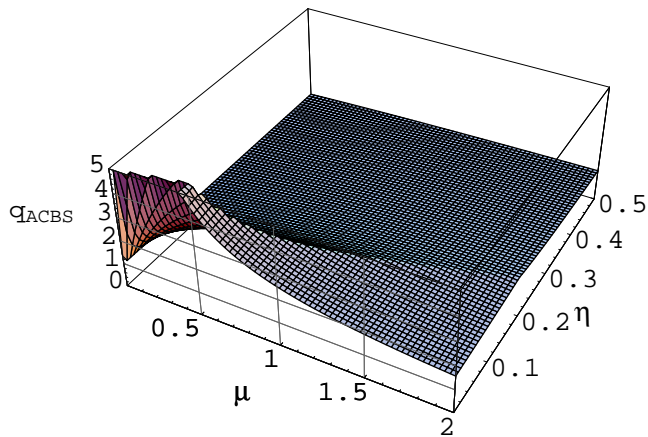,width=3.4in}}
    \caption{Performance quotient $q_{\ACBS}$ for the \ACBS.}
    \protect\label{fig:qACBS}
\end{figure}

\begin{figure}
    \centerline{\psfig{figure=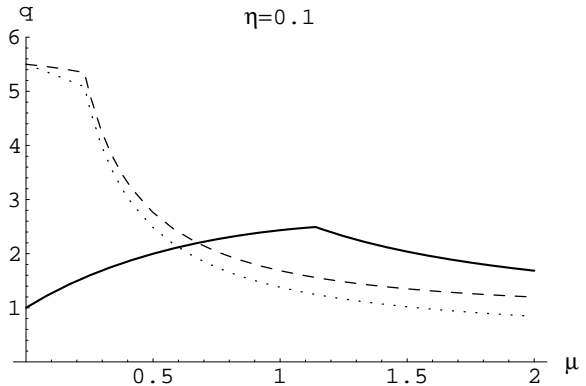,width=3.4in}}
    \caption{Performance quotient at a fixed transmissivity $\eta=0.1$ for 
    the \ACBS\ (solid), $\DCBS$ (dotted) and \CBS\ (dashed).}
    \protect\label{fig:qtfix}
\end{figure}

The performance quotient is now $q_{\ACBS}=
\frac{f_{\ACBS}}{f_{\DBS}}$.  In Fig.~\ref{fig:qACBS} we can 
see the values of this ratio as a function of the tranmissivity of 
the channel and the mean photon number.
Notice that in this case the performance quotient is larger than 
unity for all values of the mean photon number,
which means that the \ACBS\ is more efficient than \DBS. 
On the other hand, for low mean photon numbers ($\mu<1$), higher losses 
are required to achieve the same performance as the \DCBS\ (see also 
Figs.~\ref{fig:reg} and \ref{fig:frqa}).
\noindent We see that the \ACBS\ can still be substantially 
better that the \DBS. For the
values $\mu=0.1$ and $\eta=0.1$ the \ACBS\ attack provides Eve with a 
fraction of the key $q_{\ACBS}=1.3$ times bigger than the \DBS\ attack.
In Fig.~\ref{fig:qtfix} we can see the behavior of the performance
quotient as a function of the mean photon number
for a fixed value of the losses. We observe that, contrary to the 
other attacks, for a fixed transmissivity of the channel $\eta$,
 the efficiency of \ACBS\ over \DBS\ can increase with $\mu$.
For example, if the mean photon number of the previous example 
is increased to $\mu=1.1$  keeping the transmissivity in $\eta=0.1$
the efficiency factor grows to $q_{\ACBS}=2.48$.

In Fig.~\ref{fig:reg} we represent which of the studied attacks is 
most effective in the no-storage scenario for different values
of the parameters $\mu$ and $\eta$. As a summary, in Fig.~\ref{fig:frqa} 
we show the key fraction as a function of the losses for four 
different values of the mean photon and for the different attacks 
studied in this paper. For comparison, the results for 
the \PNS\ in the storage and no-storage scenarios are also plotted.
Once again, we notice that the \CBS\ and \ACBS\ provide 
real alternatives to the, at present, unfeasible \PNS\ and
the ineffective \BS.

\end{multicols}
\widetext

\begin{multicols}{2}

\section{Photon statistics}\label{sec:phstat}
The task of an eavesdropper is to acquire knowledge about the secret 
key that Alice and Bob want to share. But the important point is that 
she does this without leaving any trace in the signal that could 
indicate her presence to Alice and Bob. Eve's eavesdropping 
capabilities are therefore strongly dependent on the capabilities of 
Alice and Bob to prepare and analyze their signals. In this work we 
have assumed that Alice is not able to prepare single photon signals, 
using instead very weak coherent states. Moreover, we have 
assumed that Bob's capabilities are limited as well by 
a detection setup that gives the same outcome for single photons than 
for multiphoton signals.
In this situation Eve has only to forward signals to Bob in
such a way that expected number of non-vacuum signals is
the same as for the lossy channel, as expressed 
in Eq.\eqref{eq:p0cond}. 

In this section we will study what happens in the situations where Bob's 
capabilities are not so poor. 
In particular we consider the realistic situation in which Bob's 
analyzer is a polarizing beam splitter 
(in two possible orientations according to the basis measured) 
with a photon detector in each of its arms. We will 
assume that these detectors do not have photon number resolution, so 
that they only have two possible outcomes corresponding to
non-vacuum (`click') and vacuum (no `click') impinging signals.

When Bob and Alice bases coincide, only one of Bob's detectors can 
click and Eve remains safe. But in the case that Bob measures in a 
different basis than the encoding basis, the multiphoton part of the 
signal can lead to simultaneous clicks in both detectors.
Here is where Eve can reveal her presence depending on what attack she 
chooses.  Bob is expecting to receive a weak coherent state of a given 
amplitude (determined by the amplitude of the selected
coherent state and by the channel 
losses) and therefore an expected number of these double clicks. 
The probability of these double clicks without Eve's intervention
(or under the \BS\ attack) is given by
\begin{equation}
    p^{dc}_{\BS}=\frac{1}{2}(1-{\rm e}^{-\frac{\mu \eta}{2}})^2\mbox{,}
	\label{eq:pbsdc}
\end{equation}
where the factor $\frac{1}{2}$ accounts for the probability
that Alice and Bob  use different basis.

\begin{figure}[htbp]
    \centerline{\psfig{figure=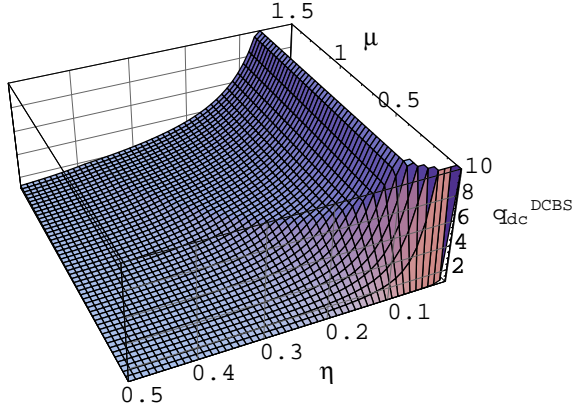,width=3in}}
    \caption{Ratio  of the \CBS\ and \BS\ double-click probabilities. }
    \protect\label{fig:ratpb}
\end{figure}

When Eve tries to eavesdrop using the \CBS\ attack 
without extra blocking (i.e $\eta > \eta_{\CBS}^{{\rm min}}$) 
 the probability of double counts is,
\begin{eqnarray}
    p^{dc}_{\CBS} & = & 
     \frac{1}{2} p_{0}(\tau)(1-{\rm e}^{-\frac{\gamma_{\tau}\mu}{2}})^2
   +\frac{1}{2}\int_{0}^{\tau}  p_{1}(t)
   (1-{\rm e}^{-\frac{\gamma_{t}\mu}{2}})^2 d t \nonumber\\ 
   &=&  \frac{1}{2}-\frac{{\rm e}^{-\mu}}{2}
   (4{\rm e}^{\frac{\mu}{2}}-1-\mu(1-\gamma_{\tau}^2)-
   2 {\rm e}^{-\frac{\gamma_{\tau}^2 \mu}{2}})\mbox{.}
    \label{eq:pcbsdc}
\end{eqnarray}
By using \eqref{eq:g2t} to express $\gamma_{\tau}$ in 
terms of the the channel losses, and taking into account the blocking 
probability $p^{{\rm block}}_{\DCBS}$ \eqref{eq:pblock} for 
$\eta<\eta_{\CBS}^{{\rm min}}$ we plot in Fig.~\ref{fig:ratpb} the 
ratio of both probabilities $q^{dc}_{\CBS}=\frac{p^{dc}_{\CBS}}{p^{dc}_{\BS}}$
as a function of the mean photon number and channel transmissivity.

\begin{figure}[!htbp]
    \centerline{\psfig{figure=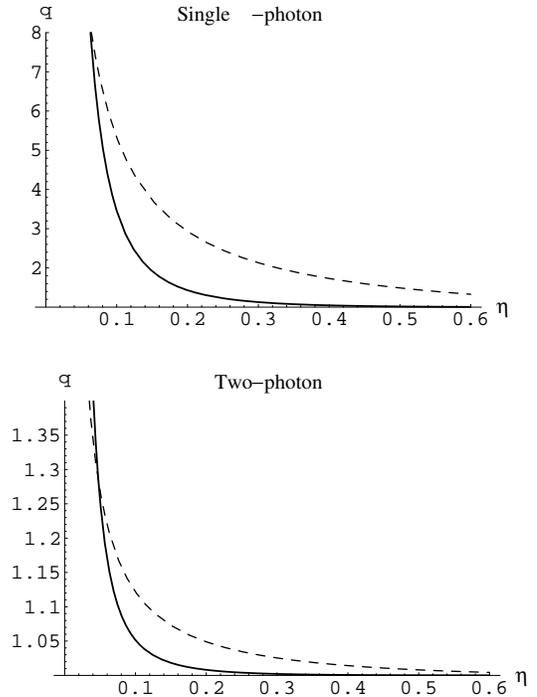,width=3in}}
    \caption{Double-click ratio (solid) and performance quotient (Dashed) at a fix $\mu=0.1$ 
    for \CBS\ and $\DCBS$ (top) and \ACBS (bottom)}
    \protect\label{fig:qdcq}
\end{figure}

We see that the \CBS\ increases the probability of double-clicks 
relative to the lossy channel. The reason for this increase is that
the \CBS\ only takes one photon out while several signal photons might be 
lost in the lossy channel.  For increasing transmissivities the differences 
tend to vanish. So, for example $\mu=\{0.1,0.01\}$ and 
$\eta=0.3$ lead to  $q_{dc}=\{1.1,1.02\}$. But for higher losses 
the ratio can be quite high, 
e.g for $\mu=\{0.1,0.01\}$ and $\eta=0.1$ 
$q_{dc}=\{3.45,1.2\}$. Inclusion of the effect of dark counts in
Bob's detectors in  $p^{dc}_{\CBS}$ shows only a slight 
reduction of the discrepancies for reasonable dark count 
probabilities ($p_{\mbox{D}} \sim 10^{-5}$). Even though this is a 
definite handicap of the \CBS\, it might be of relative importance in 
practice. The probability of double counts is very small  
($p^{dc}\sim\frac{1}{8}(\eta\mu)^2$)
while the statistical fluctuations are large,
therefore the number of transmitted signals needed to appreciate  
Eve's intervention might be exorbitant. As shown in 
Fig.~\ref{fig:qdcq}, for some parameter regimes one can 
find a good compromise between the probability of success
and the double-count rate.

\begin{figure}[!htbp]
    \centerline{\psfig{figure=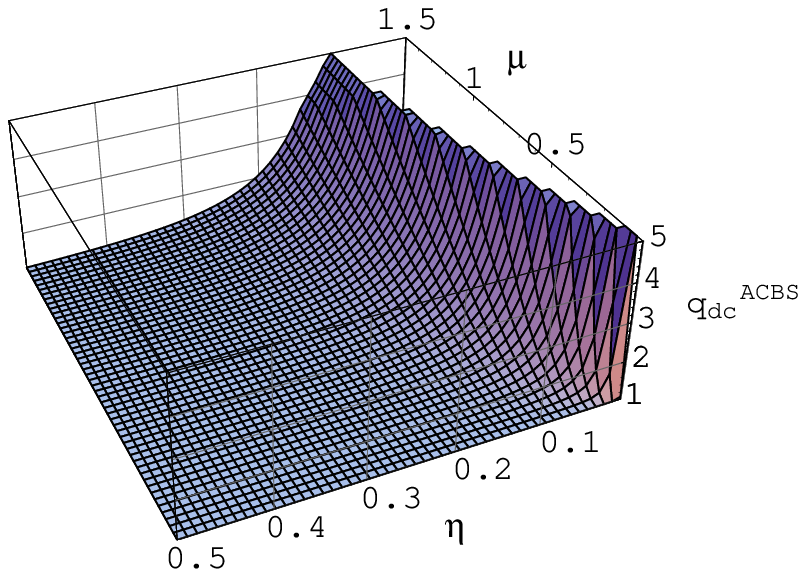,width=3in}}
    \caption{Ratio  of the \ACBS\ and \BS\ double-click probabilities. }
    \protect\label{fig:ratpb2}
\end{figure}

Moreover, as we shall see next, the disparity in the 
number of double-clicks can be further decreased if Eve uses the 
two-photon splitting adopted for the non-storage 
conditions (\ACBS). The double-click probability in this case is
\begin{eqnarray}
    p^{dc}_{\ACBS}&=&1-{\rm e}^{-\mu}\left(8{\rm e}^{\frac{\mu}{2}}-1-
        \mu(1-\gamma_{\tau}^2)-
       \frac{1}{2}\mu^2(1-\gamma_{\tau}^2)^2 +\right. \nonumber\\
       &  &\left. -2{\rm e}^{-\frac{\gamma_{\tau}^2 \mu}{2}}
       (3+\mu(1-\gamma_{\tau}^2))\right)\mbox{.}
    \label{eq:pDCBSdc}
\end{eqnarray}
In Fig.~\ref{fig:ratpb2} we can see the ratio between this probability 
and the corresponding probability for the lossy channel
$q^{dc}_{\ACBS}=\frac{p^{dc}_{\ACBS}}{p^{dc}_{\BS}}$, taking into 
account the blocking for $\eta < \eta_{\ACBS}^{{\rm min}}$. In 
Fig.~\ref{fig:qdcq} the same quantity is compared to the performance quotient.

\noindent We see that for the $\ACBS$ attack the double click 
probability takes nearly the same values as for the lossy channel
($\mu=0.1$ and $\eta=0.1$ give $q_{dc}=1.05$) while 
still providing larger key fractions than the \BS\ attack.

\section{Mixed strategies}\label{sec:mixstrat}

In the previous sections Eve's attacks were characterized by only 
one parameter $\gamma_{\tau}$. Once the type of attack was chosen (\CBS, 
\ACBS) she could only vary the coupling time $\tau$ to tune her attack.
In this section we will study the situation in which Eve can use 
mixed strategies, i.e. for each signal she can choose a different 
coupling time. In this way, Eve has the freedom to choose a 
probability distribution $\{p_{i}\}$ according to which she will 
apply a coupling time $\tau_{i}$ (i.e. $\gamma_{\tau_{i}}$). 
A mixed strategy could in principle lead to a lowering of
the double-click probability without sacrificing  to much probability 
of success. 
Unfortunately we will see that this is not the case for the 
\CBS\ and \ACBS. 

We start by considering the \CBS\ attack. In this case 
the number of non-vacuum signals received by Bob \eqref{eq:p0g}
is linearly dependent on  $\gamma_{\tau_{i}}^2$, therefore the mixed 
strategy has to be such that the average  value of 
$\gamma_{\tau_{i}}^2$ is equal to the pure strategy value fixed in 
Eq.~\eqref{eq:g2t}, i.e. 

\begin{eqnarray}
    & &\bar{P}^{{\rm B}}_{\CBS}[0]  =  \sum_{i} p_{i}  
    P^{{\rm B}}_{\CBS}[0]_{i} \nonumber \\
	& &=\sum_{i} p_{i} {\rm e}^{-\mu}(1+\mu(1-\gamma_{\tau_{i}}^2)) = P^{{\rm B}}_{\CBS}[0]
    \label{eq:p0condmix}  \\
    \longrightarrow &   & \sum_{i} p_{i} \gamma_{\tau_{i}}^2=\gamma_{\tau}^2. 
    \label{eq:g2tmixed}
\end{eqnarray}

Now we want to know what is the behavior of the probability of 
success and of the double-clicks  when using mixed 
strategies. From Jensen's inequality \cite{jensen} we know that
\begin{equation}
	f''(x)\stackrel{\geq}{<} 0 \Longleftrightarrow 
	\sum_{i} p_{i} f(x_{i}) \stackrel{\geq}{<} f(\sum_{i} p_{i} x_{i}).
	\label{eq:jensen}
\end{equation}
Since Eq.~\eqref{eq:g2tmixed} assures that Bob gets the expected 
number of clicks we find, by using Eq.~\eqref{eq:psucccbs}, that the 
mixed strategy always leads to a smaller probability of success than 
a pure one
\begin{eqnarray}
 \bar{p}^{{\rm succ}}_{\CBS}&=& \sum_{i} p_{i} 
	p^{{\rm succ}}_{\CBS}(\gamma_{\tau_{i}}^2)  \mbox{ and } 
	\frac{d^2 p^{{\rm succ}}_{\CBS}}{d (\gamma_{\tau}^2)^2}<0 \nonumber\\
& & 	\Longleftrightarrow 
	\bar{p}^{{\rm succ}}_{\CBS} < p^{{\rm succ}}_{\CBS}(\gamma_{\tau}^2)
	\label{eq:jpsucc}\mbox{.}
\end{eqnarray}
Similarly one can easily see that the probability of 
double-clicks in the \CBS\ $p^{dc}_{\CBS}$ \eqref{eq:pcbsdc} 
increases when mixing strategies. Since we already saw that the 
double-click probability is higher than for the lossy channel we 
arrive to the conclusion that using mixed strategies does 
not offer any advantages to the \CBS\  attack.

The same happens in the two-photon conditional beam splitting (\ACBS).
To prove this, one has to write the probability of success and of 
double-clicks as a function of Bob's probability of receiving a 
non-vacuum signal, and check for its concavity/covexity.

\section{CBS with finite beam splitters}\label{sec:CBSF}

In this section we will see  how to implement the \CBS\ attack with 
finite reflectivity beam splitters. The beam splitters have 
now a finite reflectivity $|r|^2$ (and transmissivity
$|t|^{2}=1-|r|^{2}$). Without any loss of generality 
we will assume that $r$ and $t$ for the signal are real. We also 
assume that they are independent of polarization.

The initial state can be written as $\ket{\phi_{0}}=\ket{\alpha;\beta}\ket{0;0}$ 
where the first ket is Alice's coherent state and the second is Eve's 
mode. After the first beam splitter the state of the system is 
\begin{equation}
   \ket{\phi_{1}}=\ket{t \alpha;t \beta}\ket{r \alpha;r \beta}\mbox{.}
    \label{eq:1bs}
\end{equation}
Notice that Eve's and the signal modes are still in a separable state. 
This means that Bob's state will be the same independently of what Eve does 
to her modes.  For coherent input states,
Bob's state will depend only on the number of beam splitters put in 
by Eve during the attack.  Eqs.\eqref{eq:nojump2} and 
\eqref{eq:jump2} reflect this situation in the infinitesimal beam 
splitting case.
This simplified the calculations leading to the results in this paper,
but the quantum jump method described in Sec.~\ref{sec:CBS}
can be used to describe the conditional beam splitting 
for any kind of input  as long as Eve's 
modes are in the vacuum state initially \cite{calsam01}.

After the first beam splitter Eve will try to detect the presence of 
photons in her modes. With probability 
$p_{1}^0=p_{c}^0(r^2 (|\alpha|^2+|\beta|^2))={\rm e}^{-\mu r^2}$ 
she will detect no photon.
If this happens she will split the signal with a second beam splitter.
She will repeat this process until she detects some photons in her modes.
After the $m^{th}$ beamsplitting the state of the system is
\begin{equation}
    \ket{\phi_{m}}=\ket{t^m \alpha;t^m \beta}\ket{r t^{m-1}\alpha;r 
    t^{m-1}\beta}\mbox{.}
    \label{eq:nbs}
\end{equation}
To keep the notation simple, we are omitting here the tested vaccum modes 
of earlier beam splitters.
The total probability of reaching this state is,
\begin{eqnarray}
    p_{m}&=&\prod_{i=1}^{m-1}p_{c}^0(\mu r^2 t^{2(m-1)})=\prod_{i=1}^{m-1}
    {\rm e}^{-\mu r^2 t^{2(i-1)}}= \nonumber\\
    & = & {\rm e}^{-\mu r^2 \sum_{i=1}^{m-1}t^{2(i-1)}}=
    {\rm e}^{-\mu (1-t^{2(m-1)})}. 
    \label{eq:prnbs}
\end{eqnarray}
Therefore the total probability that Eve detects some signal after the $m^{th}$ 
beam splitter is
\begin{eqnarray}
	p_{m}^{\neg 0}&=& p_{m} p_{c}^{\neg 0}(\mu r^2 t^{2(m-1)}) \nonumber\\
	&=&{\rm e}^{-\mu (1-t^{2(m-1)})} (1-{\rm e}^{-\mu r^{2} t^{2(m-1)}})\mbox{.}
	\label{eq:pmneg} 
\end{eqnarray}
If this event occurs, the splitting stops and the signal is sent
to Bob through the lossless channel. Otherwise Eve keeps on adding 
beam splitters until she reaches a maximum number $N$ of attempts.
After the conditional beam splitting attack 
with finite beam splitters (\CBSF ), Bob  will therefore receive a coherent state 
$\ket{\phi^{bob}_{m}}=\ket{t^{m} \alpha}$ 
with probability $p_{m}^{\neg 0}$ for $m=1\ldots N-1$ or 
$p_{N}$ for $m=N$. 
From here one can calculate the probability of vacuum signals 
arriving at Bob's site in the \CBSF\
\begin{eqnarray}
    P^{{\rm B}}_{\CBSF}[0] & = & p_{N} p_{c}^0(\mu t^{2N})+ 
	\sum_{m=1}^{N-1}p_{m}^{\neg 0} 
    p_{c}^0(\mu t^{2m})  \nonumber\\ 
   &=& {\rm e}^{-\mu}\left(1-N+\sum_{n=0}^{N-1}
   {\rm e}^{\mu r^2 t^{2n}}\right)\mbox{.}   \label{eq:p0gf}
\end{eqnarray}
The probability of success of the \CBSF\ attack is,
\begin{eqnarray}
	p_{\CBSF}^{{\rm succ}} & = & \sum_{m=1}^{N}p_{m}^{\neg 0} 
    p_{c}^{\neg 0}(\mu t^{2m})= \sum_{m=1}^{N}p_{m} 
    (1-{\rm e}^{-\mu t^{2m}}) \nonumber  \\
	 & = & 1+{\rm e}^{-\mu}\left(N-{\rm e}^{\mu t^{2N}}-
	 \sum_{n=0}^{N-1}{\rm e}^{\mu r^2 t^{2n}}\right)
	\label{eq:pscbsf}
\end{eqnarray}
Notice that Eqs.~\eqref{eq:pmneg} and \eqref{eq:prnbs} and the derived 
quantities are in agreement up to first order in $r^2$ with the corresponding 
probabilities for the infinitesimal \CBS\ 
 derived with the quantum jump method, 
i.e. Eqs. \eqref{eq:prbnophoton} and \eqref{eq:prbjump}
(with $r^2=\frac{\tau}{N} \epsilon^2$).

In order to match Bob's probability of detecting a non-vacuum signal 
 with the result he would expect from the lossy channel,
 Eve has now two free parameters:
the maximum number $N$ of beam splitters used and their 
transmission amplitude $t$. From the infinitesimal \CBS\ results 
 we know that for small reflection coefficients (i.e. $t \sim 1 $),
$t^{2N}\approx \gamma_{\tau}^2$  taking $\gamma_{\tau}^2$ from 
Eq.~\eqref{eq:g2t}.  Of course, for a finite $N$ this approximation will 
not hold when we are in or near the region  
$\eta < \eta_{\CBS}^{{\rm min}}$ which corresponds to $\tau \rightarrow 
\infty$. The reason for this is that now we are dealing with 
finite beam splitters and accordingly, for any given $N$,  we can `mimic' a
lossy channel with arbitrarily high losses. For the same reason now
we are not forced to block any signals as done for the \CBS\ 
in the high losses regime.

To get the quantitative results presented in the rest of this 
section we have to find numerically the condition on Eve's 
free parameters ($t$ and $N$) such that 
Bob's expectations on the number of non-vacuum signals are fulfilled.
For a given $N$ we use Newton's method
to find the value of $t$ for which 
  $P^{{\rm B}}_{\CBSF}[0]$ \eqref{eq:p0gf} equals to
$ P^{{\rm B}}_{\eta}[0]$ ~\eqref{eq:p0B}, taking a starting value of
 $t_{o}= \eta^{\frac{1}{2N}}$.

\begin{figure}
    \centerline{\psfig{figure=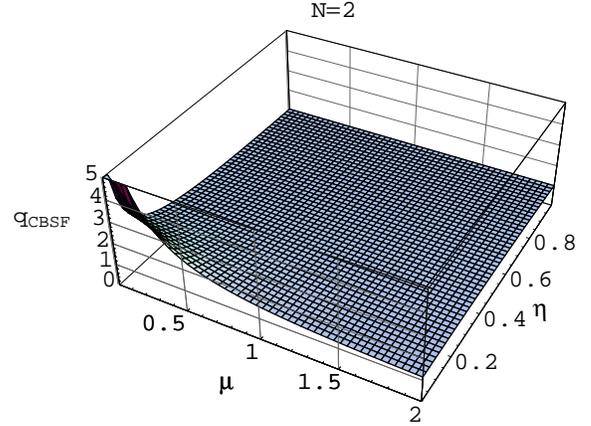,width=3in}}
    \caption{Ratio between the \CBSF\ with two beam splitters and the \BS\ 
	success probabilities.}
    \protect\label{fig:qcbsf2}
\end{figure}

\begin{figure}
    \centerline{\psfig{figure=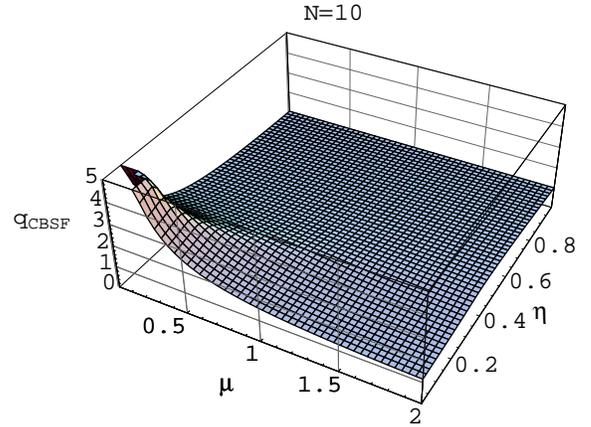,width=3in}}
    \caption{Ratio  between the \CBSF\ with ten beam splitters and  
    the \BS\ success probabilities. }
    \protect\label{fig:qcbsf10}
\end{figure}

In Fig.~\ref{fig:qcbsf2} and \ref{fig:qcbsf10} we have plotted the 
performance quotient $q_{\CBSF}=\frac{f_{\CBSF}}{f_{\BS}}$ as a function 
of $\mu$ and $\eta$ for an attack using a maximum of two and ten 
beams splitters respectively.
\noindent We notice that doing $\CBSF$ with only two beam splitters, 
Eve already obtains a much bigger fraction of the key than
with the standard beamsplitting attack,
e.g for $\mu=0.1$ and $\eta=0.1$ the key she obtains is  
$q_{\CBSF}^{N=2}=2.5$ times longer.
With ten beam splitters she almost reaches the infinitesimal
result (compare with Fig.~\ref{fig:qcbs1}).
For $\mu=0.1$ and $\eta=0.1$ the key she obtains is  
$q_{\CBSF}^{N=10}=4.7$ times longer.

\begin{figure}[!hbp]
    \centerline{\psfig{figure=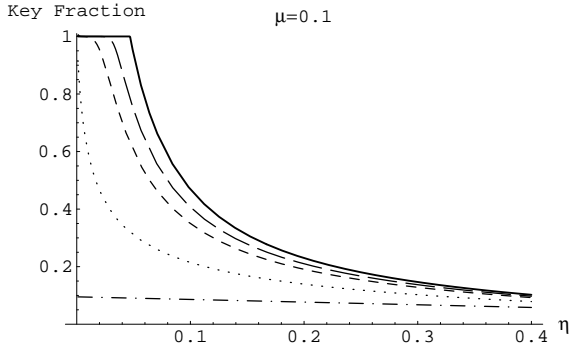,width=3in}}
    \caption{Key fraction for \CBS\ (solid), \CBSF\ with N=2 
    (Dotted), N=5 (Short-dashed), N=10 (Long-dashed) and \BS\ 
    (Dash-dotted).}
    \protect\label{fig:frcbsf}
\end{figure}

In Fig.~\ref{fig:frcbsf} we can compare the key fraction obtained 
with the \BS, \CBS\ and \CBSF\ with different numbers of beam splitters.
\noindent The number of double-clicks in Bob's detectors when he measures in 
the wrong basis  can be calculated from
\begin{equation}
    p^{dc}_{\CBSF}  =  p_{N} p_{c}^{\neg 0}(\frac{1}{2}\mu t^{2N})^2+ 
	\sum_{m=1}^{N-1}p_{m}^{\neg 0} 
    p_{c}^{\neg 0}(\frac{1}{2}\mu t^{2m})^2\label{eq:pdcf}.
\end{equation}
The ratio between the double-click 
probabilities of the \BS\ and \CBSF\ for various values of $N$, 
is plotted Fig.~\ref{fig:qdcq} 
together with the corresponding performance quotients. This figure 
shows that, depending on the parameter regime, an attack with a small number 
of finite beam splitters might be more suitable than the infinitesimal 
\CBS\, in that the number of double clicks is much closer 
to its lossy channel values.

\begin{figure}[!hbp]
    \centerline{\psfig{figure=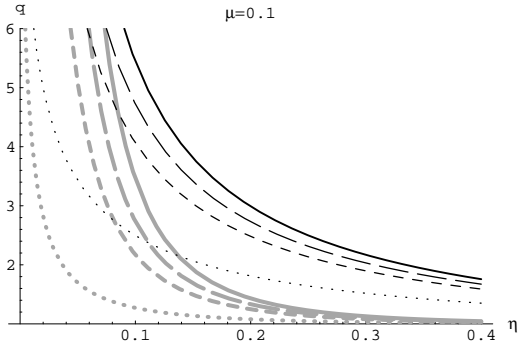,width=3in}}
    \caption{Quality factors (black) and double-click ratio (Gray) 
	for \CBS\ (solid), \CBSF\ with N=2 
    (Dotted), N=5 (Short-dashed), N=10 (Long-dashed).}
    \protect\label{fig:qdq}
\end{figure}

\section{Conclusion}\label{sec:concl}

In this paper we have introduced a novel attack on weak coherent 
pulse quantum key distribution in lossy channels. 
The conditional beam splitting takes advantage of the multiphoton 
component of the transmitted signals to extract information on the encoded 
bit. This task is accomplished optimally 
 using the photon number splitting attack.
However, the implementation of the \PNS\ attack entails a quantum non 
demolition measurement which is something  unattainable, at least at 
the single photon level, with the current technology.
Until now the technologically possible alternative to the \PNS\ 
has been the simple beam splitting attack \cite{felix01}.
For high losses (i.e. long transmission distances) the \BS\ attack 
turns out to be very ineffective. 
Here we have presented the conditional beam splitting 
attack which also requires only linear 
optical elements and is therefore feasible with present technology,
but is much more efficient than the conventional beam splitting attack.
The \CBS\ attack, thus, shortens substantially the gap
in performance between the ideal and practical eavesdropping attacks.
This is of great importance if one considers that the eavesdroppers 
success in a quantum key 
distribution attack is dictated by the technology at the moment of 
the signal transmission. In contrast with the case of classical 
cryptographic protocols, no future technologies can help 
unveil present key exchanges.

Starting from the simplest scenario in which Eve is capable of 
storing her signals until the encoding basis is announced, we
have moved to more realistic situation in which Eve has no storing 
capabilities. For this situation an adapted \CBS\ attack, based on the 
extraction of two single photons, has been 
 proven to be advantageous for some relevant parameter regimes.

Numerical results for the implementation of \CBS\
 with finite reflectivity beam splitters show that using only 
 two beam splitters
 one can easily duplicate the efficiency of the conventional beam 
 splitting, and that the infinitessimal \CBS\ results are reached 
 with the use of a few beam splitters.

The photon statistics at Bob's detectors has been 
studied and shown to be a matter of concern in this type of attack.
However, we argue that, if handled with care, this drawback 
does not disqualify the \CBS\ attacks.
We believe that further elaborations of this basic idea can lead to attacks 
which are specialized for certain parameter regimes and other protocols.

\section*{Acknowledgements}\label{ackw}
S. M. B.  thanks the Royal Society of Edinburgh and the Scottish Executive 
Education and Lifelong Learning Department for financial support.
J.C. acknowledges the Academy of Finland (project 4336)
and the European Union IST EQUIP Programme for financial support.


\begin{thebibliography}{10}
\bibitem{note1}
Naturally, this requires Eve to be able to perform a
\QND\ measurement in order to determine the presence of
a single photon without absorbing it.  This is not 
realistic with current technology.  Allowing Eve to
do this, however, serves to demonstrate the full 
potential of the CBS attack.
\bibitem{note2}
For simplicity we will not 
consider in this paper the effects of 
any imperfections in the channel used by Eve to send the photons to Bob.
In any case it is reasonable to assume that Eve can provide a channel with lower 
losses than the one used by Alice and Bob. The effect of considering 
Eve's losses is equivalent to the effect of Bob's detector efficiency.
\bibitem{note3}
Eve can use a cascade of beams splitters in 
order to distribute the photons she receives so that the 
probability of having two photon detections is infinitesimally small.
\end{thebibliography}

\end{multicols}

\end{document}